\documentstyle[editedvolume,psfig,epsf]{crckapb} 

\begin{opening}
\title{Revisiting the Open Cluster  M35}

\author{D. BARRADO Y NAVASCU\'ES}
\institute{Depto. F\'{\i}sica Te\'orica, C-XI-506,\\ 
Universidad Aut\'onoma de Madrid\\
           28049 Madrid, SPAIN}
\end{opening}

\runningtitle{Revisiting the Open Cluster M35}

\begin{document}


\begin{abstract}
 We present a study of different properties of the young open cluster M35,
including
very deep and accurate photometry, rotational and radial velocities,
and lithium abundances.

\end{abstract}

\section{Previous studies of the cluster.}

M35 is a young, rich open cluster. During the last years, different
authors have studied its properties in detail.
Sung \& Bessell  (1999) have derived a distance modulus of
(m-M)$_0$=9.60$\pm$0.10, a reddening of 
E(B--V)=0.255$\pm$0.024 and an age of $\tau$=200$_{-100}^{+200}$ Myr.
Sarrazine et al. (2000) have obtained 
(m-M)=10.16$\pm$0.01, E(B--V)=0.198$\pm$0.008 and 
 $\tau$=160$\pm40$ Myr.
On the other hand, proper motions for stars warmer than late
 F spectral type  have been measured by
McNamara \& Sekiguchi (1985) and 
the total dynamical mass has  been evaluated 
as 1600--3200 M$_\odot$ (Leonard \& Merritt 1989).
Prior to our study, about 1100 candidate members were identified, 
down to V$\le$19.5 mag.

\begin{figure}
\vspace{10.8cm}
\includegraphics{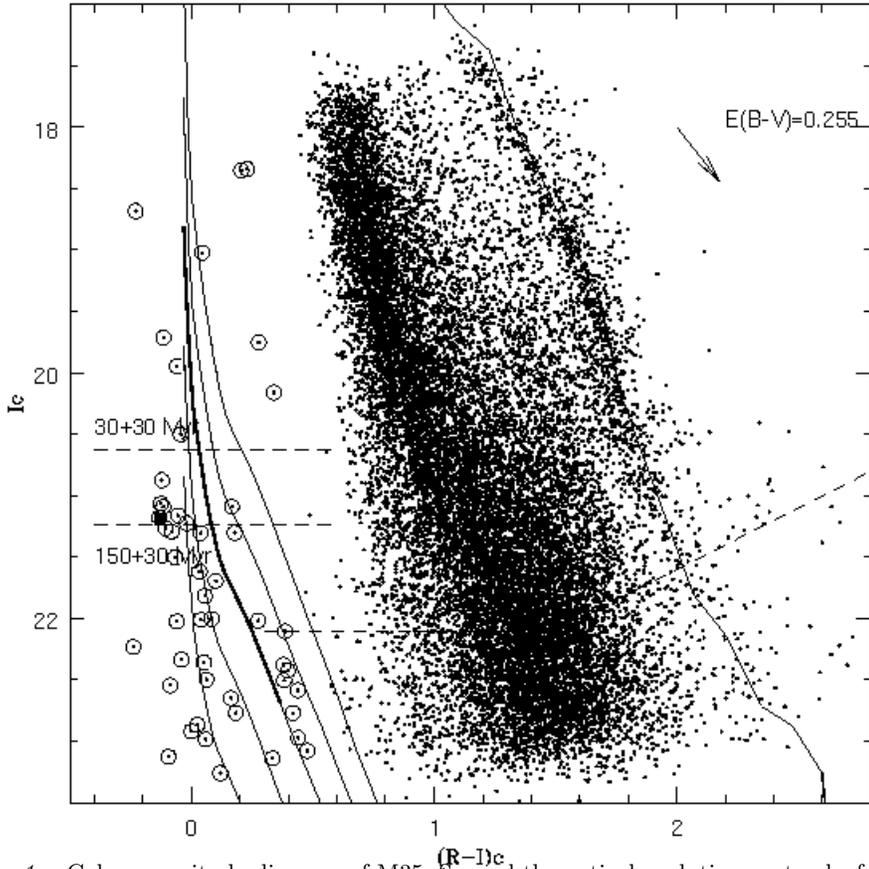}
\caption{Color-magnitude diagram of M35. Several theoretical
evolutionary tracks for WD are included.}
\end{figure}

\begin{figure}	
\vspace{-0.0cm}
\hbox{\hspace{2mm}\epsfxsize=6.0cm \epsfbox{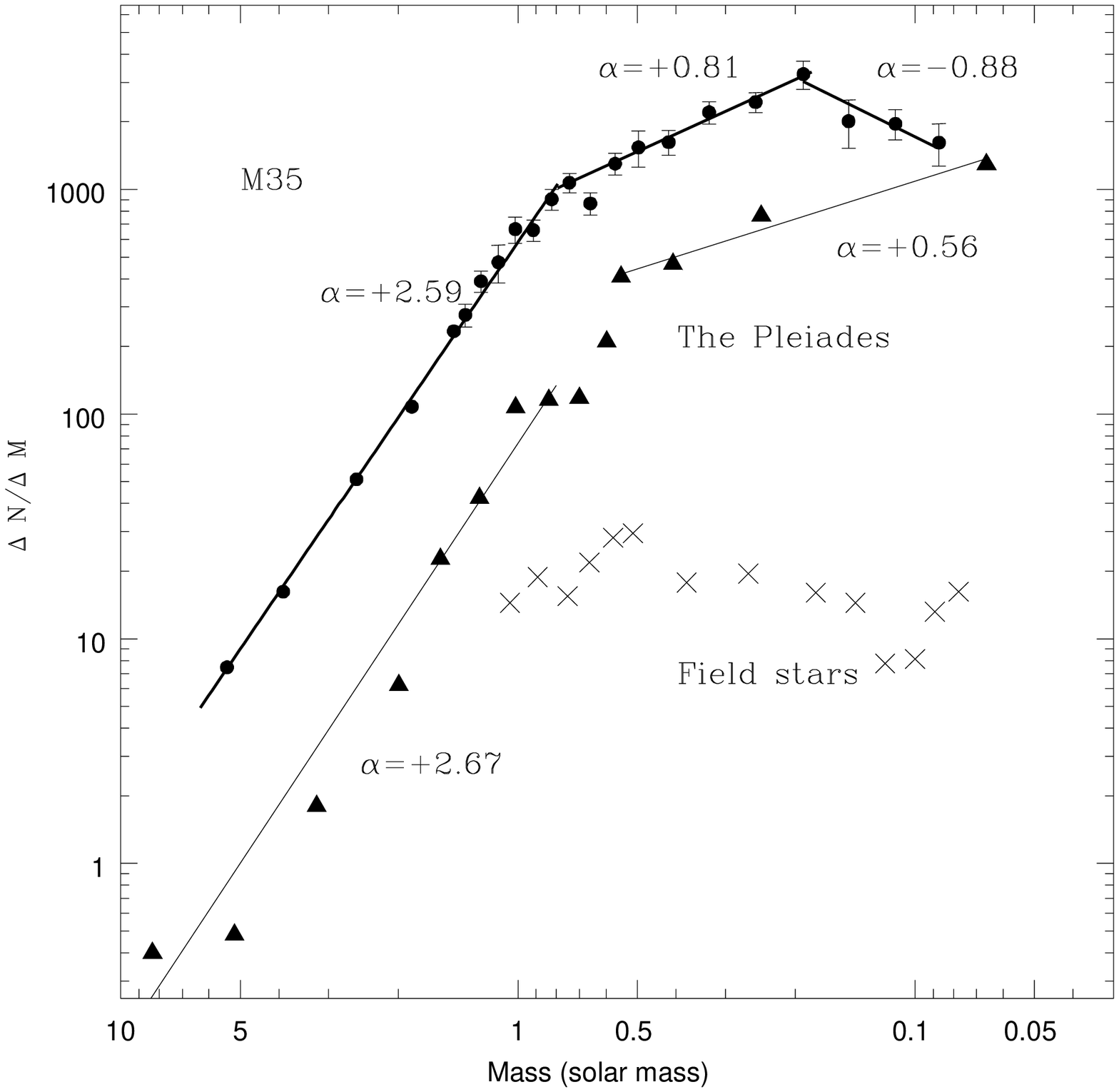}
\hspace{2.8mm}\epsfxsize=6.0cm \epsfbox{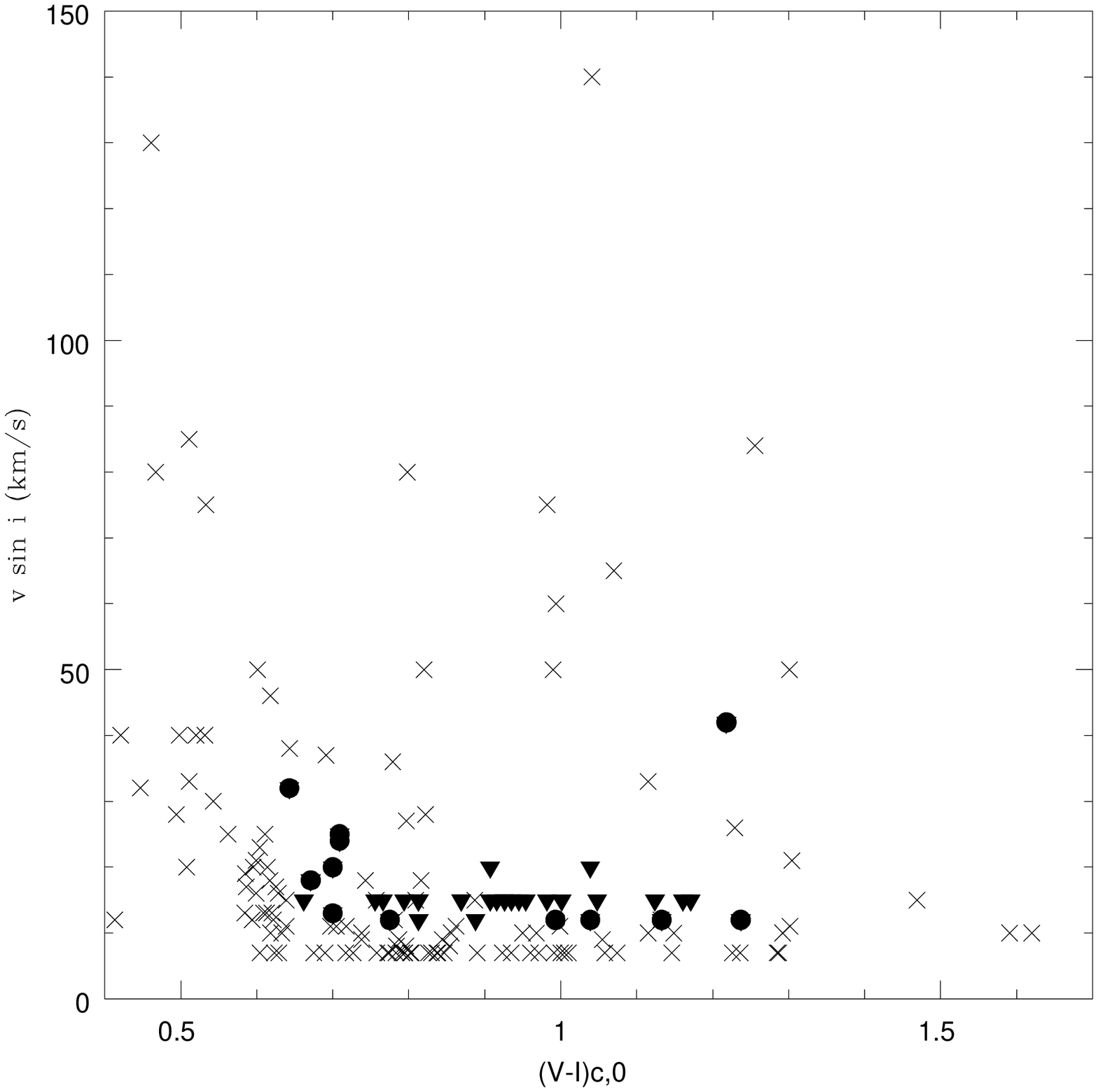}}
\vspace{-2mm}
  \caption{
{\bf a} Complete Mass Function for the stellar population 
 of M35 (solid circles). We have included a comparison with
the Pleiades (solid triangles) and field stars (crosses).  
{\bf b} Rotational velocity distribution for M35 stars (solid symbols)
Pleiades data are displayed as crosses.
}
\end{figure}

\section{A new vision of M35.}

We have carried out several observing runs, both photometric --V(RI)$_C$
 filters-- and
spectroscopic (R$\sim$20,000),  with several goals:

(i) By using color-magnitude diagrams, we have identified
new low mass candidate members of the cluster, down to I$_C$=23.5 mag,
including 65 brown dwarf candidates and another $\sim$1700 stellar
 candidate  members. The photometry comes from KPNO 4m telescope and
Canadian-France-Hawaii telescope. Figure 1 shows one of our 
color-magnitude diagrams.

(ii) The new extended population of tentative members of the cluster
has been used to derive a accurate Luminosity Function  and Mass Function
(MF) 
for the totality stellar population of the clusters
(6--0.08 M$_\odot$).   Figure 2a depicts the MF and a
comparison with the Pleiades (80--100 Myr, 
data from Hambly et al. 1999)
and field stars (Gould et al. 1997).
The main characteristics of the M35 MF are its accuracy (due to its richness)
and the three different indices $\alpha$
 depending on the mass range (see the figure).
These discontinuities in the MF could indicate that there were several
mechanisms working when the initial cloud fragmented and collapsed.
In addition, we have estimated the total mass of the core of the cluster, 
1600 M$_\odot$.

(iii) We have identified a population of white dwarfs (WD).
The LF of WD and the comparison with theoretical isochrones
suggest that the clusters is 180 Myr (see Figure 1).

(iv)  The high resolution multi-fiber spectroscopy (WIYN/HYDRA)
 for 76 photometric
candidate members were used to select a population of {\it bona fide}
members (39 stars of late F, G and early K spectral types). Another
13 objects present variable radial velocities in a two days time span.
 (Three  are SB2 binaries.)
We consider them as possible members of the cluster.

(v) Cross-correlation techniques were used to derive rotational 
velocities for these 76 stars. The distribution of {\it v sini}
versus the (V-I)$_C$ color index is shown in Figure 2b (solid triangles
-upper limits- and circles).
M35 lacks fast rotators in this color range, in opposition to
what happens in the Pleiades (crosses). This distribution confirms that
M35 is older than the Pleiades.

(vi) Our free spectral range (6440--6850 \AA) contains a dozen
iron lines. We have derived the metal content for the brightest
stars in the sample using spectral  synthesis and 
model atmospheres (Kurucz 1992).
The M35 metallicity is [Fe/H]=--0.21$\pm$0.10, whereas 
the Pleiades value is $+$0.01 using the same procedure.

(vii) We have measured lithium 6707.8 \AA{ }
 equivalent widths. Lithium abundances were derived using curves
of growth (Soderblom et al. 1993).
 Figure 3 displays the M35 lithium abundances versus 
effective temperatures (solid symbols), together with a comparison 
with Pleiades stars (crosses). Clearly, the lithium spread
present in the Pleiades  (Butler et al. 1987; Soderblom et al. 1993) 
does not appear in M35. For a given 
temperature, M35 abundances tend to be smaller. 
Our high quality M35 database of lithium abundances and rotational velocities
is perfectly suited to be used as a laboratory
to test theoretical models dealing with the lithium depletion phenomenon.

Additional details can be found in Barrado y Navascu\'es
et al. (1999) and Barrado y Navascu\'es
et al. (2000a,b).

\begin{figure}
\vspace{8.5cm}
\includegraphics{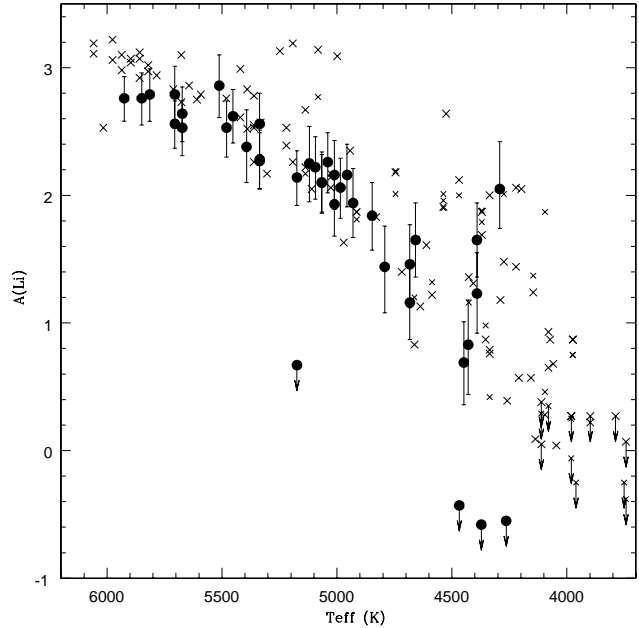}
\caption{The effective temperature -- lithium abundance plane.
M35 members are shown as solid symbols, whereas Pleiades stars appear
as crosses.}
\end{figure}

\acknowledgements
 This work has been partially suported by 
Spanish {\it ``Plan Nacional del Espacio''}, under grant ESP98--1339-CO2.

\end{document}